# Triangular spectral phase tailoring for the generation of high-quality picosecond pulse trains

U. Andral, J. Fatome, B. Kibler and C. Finot

*Abstract*— We theoretically and experimentally demonstrate a new electro-optic linear approach to generate high repetition-rate picosecond pulse trains. This simple cavity-free method is based on a temporal sinusoidal phase modulation combined with a triangular spectral phase processing. Experimental results validate the concept at repetition rates ranging from 10 GHz up to 40 GHz with the generation of background-free pulse trains made of nearly Gaussian Fourier-transform limited pulses.

## I. Introduction

The generation of high quality pulse trains at repetition rates of several tens of GHz remains a crucial step for optical telecommunications, optical sampling or component testing applications. Unfortunately, the current bandwidth limitations of optoelectronic devices do not allow the direct generation of well-defined optical pulse trains with low duty cycles. Actively mode-locked fiber lasers were found to be an efficient way to overcome this limitation [1] but they remain an onerous and complex option. As an alternative approach to conventional mode-locking operation, several cavity-free operation techniques have been suggested, such as the nonlinear reshaping of a sinusoidal beating into a train of well-defined, pedestal-free pulses [2]. For example, taking advantage of the distributed interaction of the anomalous dispersion of an optical fiber with its Kerr nonlinearity, multiple four-wave mixing has enabled the generation of Fourier-transform limited pulses at repetition rates exceeding 1 THz [3]. The concatenation of dispersion-managed segments of fibers [4] is another solution that may benefit from the nonlinearity of optical fibers.

An alternative set of linear solutions is based on a direct temporal phase modulation that is then converted into an intensity modulation. Such a conversion can be achieved using two different schemes. One relies on the use of a frequency shifted optical bandpass filter [5, 6]. This simple technic is efficient to produce ultrashort pulses but is intrinsically highly dissipative. A second well-known approach is based on a dispersive element that imprints a spectral quadratic phase. Picosecond pulses at repetition rates of several tens of GHz have been successfully demonstrated [7, 8]. Taking advantage of the space/time duality of dispersion, an analogy with a lenticular lens [9] or with a Fresnel diffraction pattern by a pure phase grating can be drawn. However, this approach suffers from a limited extinction ratio or from the presence of detrimental temporal sidelobes [10], so that a considerable part of the energy lies outside the main pulses. In order to reject these residual background and tails, various architectures have been proposed, such as the use of a broadband optical carrier [11], of a nonlinear optical loop mirror [12] or of an additional amplitude modulation with optical bandpass spectral filtering [13, 14].

We introduce here theoretically and experimentally an alternative scheme where the quadratic spectral phase is replaced by a triangular one. With such a specific phase processing, Fourier-transform limited structures are obtained. Experimental validation carried at repetition rates between 10 and 40 GHz confirm that high-quality close-to-Gaussian pulse trains can be achieved with an excellent extinction ratio and with a duty cycle below 1/4, in full agreement with our numerical simulations and theoretical predictions.

## II. Principle of operation

In order to illustrate our approach, let us first consider a continuous optical wave with an amplitude $\psi_0$ and a carrier frequency $f_c$, $\Psi(t) = \psi_0\, \psi(t)\, e^{i 2\pi f_c t}$ which phase is temporally modulated by a sinusoidal waveform:

$$\psi(t) = e^{i A_m \cos(2\pi f_m t)}, \qquad (1)$$

where $A_m$ is the amplitude of the phase modulation and $f_m$ its frequency. We have reported in Fig. 1 the resulting phase and intensity profiles both in the temporal and spectral domains for $A_m = 1.1$ rad. The temporal sinusoidal phase leads to a set of spectral lines that are equally spaced by $f_m$ and which amplitude can be expressed using a Jacobi-Anger expansion [15]:

$$\psi(t) = \sum_{n=-\infty}^{\infty} i^n\, J_n(A_m)\, e^{i 2\pi n f_m t} \qquad (2)$$

Therefore, the $n^{th}$ spectral component has an intensity proportional to $J_n^2$, with $J_n$ being the Bessel function of the first

This work is financially supported by PARI PHOTCOM Région Bourgogne, by the Institut Universitaire de France, by FEDER-FSE Bourgogne 2014/2020 and by the the french "Investissements d'Avenir" program, project (Labex Action ANR-11-LABX-01-01). The research work has benefited from the PICASSO experimental platform of the University of Burgundy.

U.A., J.F., B.K. and C.F. are with the Laboratoire Interdisciplinaire Carnot de Bourgogne, UMR 6303 CNRS-Université Bourgogne-Franche-Comté, 9 av. A. Savary, 21078 Dijon cedex, France (christophe.finot@u-bourgogne.fr).

kind of order *n*. An essential point that has not been stressed and exploited so far is the existence of a phase shift of π/2 between each spectral component (Fig. 1(b1)). Consequently, these equally spaced spectral phase shifts can also be described by a triangular spectral phase profile. Involving a purely dispersive element, i.e. applying a quadratic spectral phase, one can partly compensate for this initial phase (triangles on panel (b1)), leading to the emergence of temporally localized pulse structures with a period $T_0 = 1 / f_m$. Nevertheless, since the spectral phase is not perfectly canceled, a deleterious residual background and a poor extinction ratio can be observed on the resulting intensity profile (panels a2 and a3, dashed grey line) as well as an uncompensated residual temporal phase. However, with the progress of linear shaping, one can now process a line-by-line spectral phase profile using liquid-crystal modulators [16] or fiber Bragg gratings [17]. It then becomes feasible to imprint an exact triangular spectral phase profile of opposite sign. Therefore, for $A_m < 2.40$, a flat spectral phase can be achieved, leading to a Fourier-transform limited waveform $\psi'(t)$ that can be expressed as:

$$\psi'(t) = J_0(A_m) + 2\sum_{n=1}^{\infty} J_n(A_m) \cos(2\pi n f_m t) \quad (3)$$

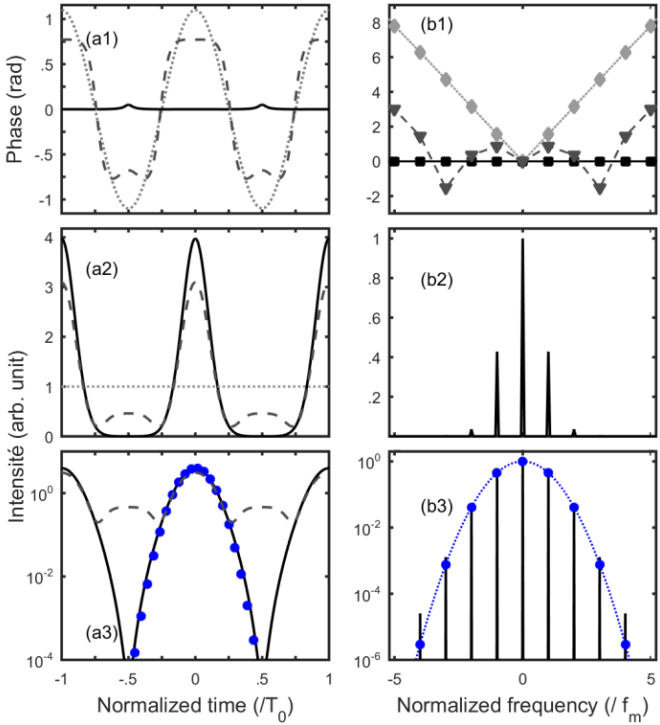

Fig. 1. (a) Temporal and (b) spectral properties of the waveform after temporal phase modulation (light grey dotted line), after spectral shaping with a triangular phase (black solid line) or with a quadratic phase (dashed grey line). The phase profiles and the intensity profiles on a linear or logarithmic scale are plotted on panels (1), (2) and (3), respectively. The blue full circles correspond to a fit by a Fourier-transform Gaussian pulse. Results obtained for $A_m = 1.1$ rad.

As a consequence, in the temporal domain, the initial phase modulation is converted into a pure intensity modulation leading to a train of well-separated ultrashort pulses at the repetition rate $f_m$ (Fig. 1(a2) and (a3)). It should be also noted that for $A_m = 1.1$ rad, the resulting intensity profile can be closely adjusted by a Fourier-transform limited Gaussian shape (see blue full circles in Fig. 1(a3) and (b3)). Finally, the resulting pulse train has a duty-cycle of 1/5 and is characterized by an excellent extinction ratio (>30 dB), well beyond the performance which can be achieved with a usual group delay dispersion circuits imprinting a quadratic spectral phase.

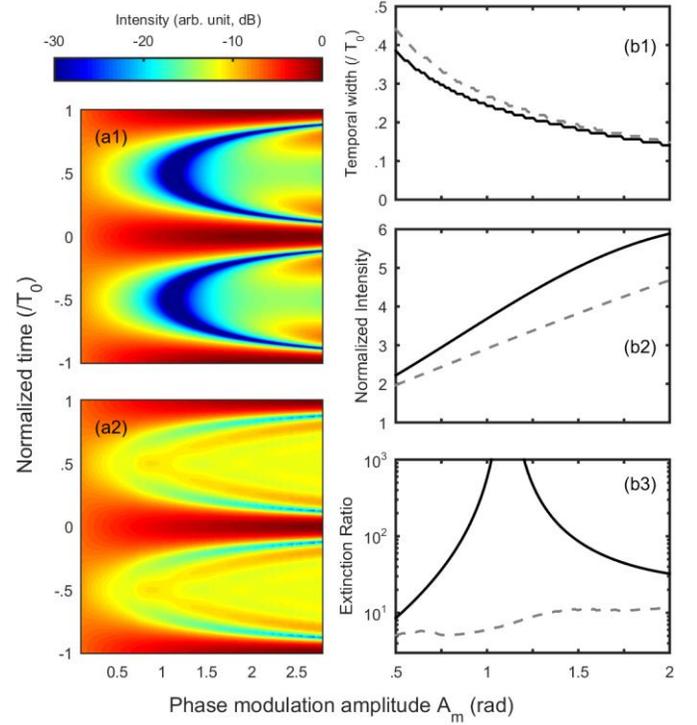

Fig. 2. Evolution of the temporal intensity properties according to the sinusoidal phase modulation amplitude. (a) Evolution of the temporal intensity profile for a triangular or parabolic spectral phase compensation, panel 1 and 2 respectively. (b) Evolution of the normalized full width at half maximum (fwhm) temporal duration of the resulting pulses, their intensity peak as well as the corresponding extinction ratio, panel 1, 2 and 3 respectively. Results obtained with a triangular spectral phase compensation (black solid line) are compared with the results from a quadratic spectral phase compensation (grey dashed line).

Figure 2 illustrates the influence of the phase modulation amplitude $A_m$ on the resulting waveform and according to the spectral compensation scheme. Higher $A_m$ is, shorter are the generated pulses with an increasing peak power. Whereas triangular or parabolic spectral phase leads to rather similar results in terms of temporal duration (panel b1), the resulting peak power (panel b2) is enhanced with the triangular scheme due to a reduced amount of energy remaining in the residual sidelobes or background. Therefore, the extinction ratio is dramatically improved for a triangular phase shaping as can be noticed in Fig. 2(a). While quadratic spectral compensation cannot enable extinction ratio well above 10, results are improved by more than one order of magnitude when triangular spectral phase is applied: for $A_m$ between 1 rad and 1.25 rad, extinction ratios above 20 dB can be expected with an optimum value obtained for $A_m = 1.1$ rad. However, for higher modulation depth of the phase, a residual background or sidelobes appear so that it is impossible to achieve outstanding extinction ratios with duty cycles above 1/6.

## III. EXPERIMENTAL SETUP AND RESULTS

The experimental setup is sketched on Fig. 3 and is based on devices that are commercially available and typical of the telecommunication industry. A continuous wave laser at 1550 nm is first temporally phase modulated using a Lithium Niobate electro-optic device driven by an amplified sinusoidal electrical signal. A linear spectral shaper (Finisar Waveshaper) based on liquid crystal on silicon technology [18] is then used to apply the suitable triangular spectral phase under test. Note that for repetition rates above 10 GHz, it becomes also possible to apply simple discrete spectral phase shifts of π/2 between two successive components. In order to ensure enhanced environmental stability, polarization-maintaining components have been used. Moreover, since the principle of operation is purely linear, no Erbium doped fiber amplifier is required, thus limiting the source of detrimental noise. The reshaping process is here quite energy efficient, since the optical losses are restricted to the insertion losses of the phase modulator and of the spectral shaper. The resulting signal is directly recorded by means of a high-speed optical sampling oscilloscope (1 ps resolution) and with a high-resolution optical spectrum analyzer (5 MHz of resolution).

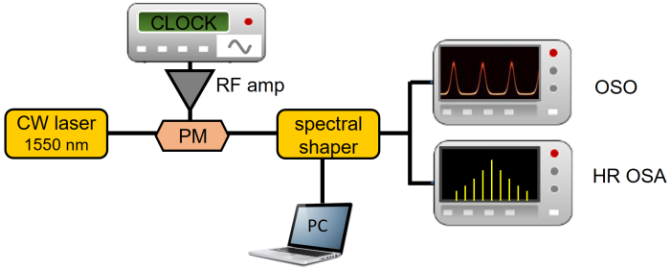

Fig. 3. Experimental setup. CW: Continuous Wave ; PM: Phase Modulator ; OSO: Optical Sampling Oscilloscope ; HR OSA: High-Resolution Optical Spectrum Analyzer

Experimental evolution of the temporal intensity profile according to the amplitude of the phase modulation is summarized in Fig. 4(a). These results obtained for $f_m = 10$ GHz confirm that the achievement of a significant extinction ratio requires to carefully choose $A_m$, typically between 1 and 1.25 rad. The experimental trends are similar to the numerical predictions depicted in Fig. 2(a1): for low modulations, a continuous background remains between two consecutive pulses whereas excessive modulations induce unwanted bounces. Positions of the minima of the temporal intensity profile are also fully in-line with the numerical simulations (white dotted line). For an optimum phase modulation depth ($A_m = 1.1$ rad), the pulses are characterized by a fwhm duration of 21 ps and the resulting temporal intensity profile (Fig. 4(b)) matches perfectly the analytical results of Eq. (3) with a summation truncated to the first three harmonic terms.

Details of the pulse train obtained for $f_m = 20$ GHz and for the optimum phase modulation $A_m = 1.1$ rad, are provided in Fig. 5. Once again, pulses with an excellent extinction ratio (> 20 dB) are recorded with a fwhm temporal duration of 12 ps in line with the numerical predictions. Figure 5(a) stresses the high level of stability of the pulse train with negligible temporal or amplitude jitter. Indeed, in contrast to nonlinear reshaping schemes that may suffer from severe Brillouin backscattering in fiber segments under test, it is not here required to involve additional phase modulation that may turn into detrimental jitters [19]. The temporal intensity profile of the pulse can be closely approximated by a Gaussian waveform (panel b) and its spectrum corresponds to the Fourier-transform limit (dotted blue line). High resolution optical spectra shown in panels (c) outline the quality of the resulting signal with an optical signal to noise ratio above 60 dB.

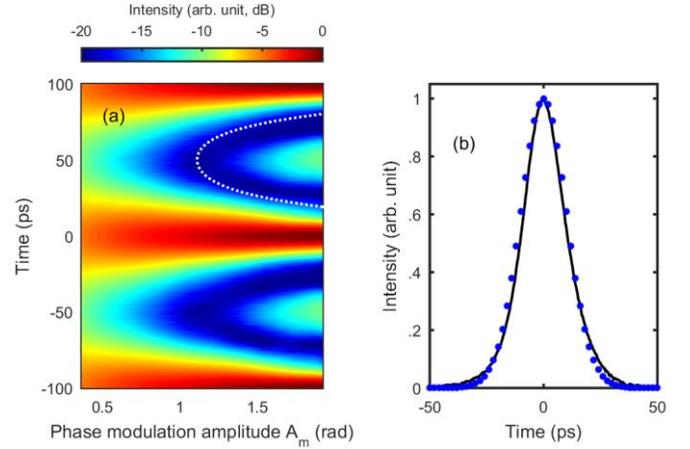

Fig. 4. (a) Evolution of the temporal intensity profile obtained after triangular phase compensation according to the sinusoidal phase modulation amplitude. The white dotted line corresponds to the position of the minima predicted numerically in Fig. 2(a1). (b) Details of the pulse structure obtained for $A_m = 1.1$ rad. The experimental results (solid black line) are compared with the analytical predictions based on Eq. (3) taking into account only the first three harmonics (blue full circles). All of the results are recorded for $f_m = 10$ GHz.

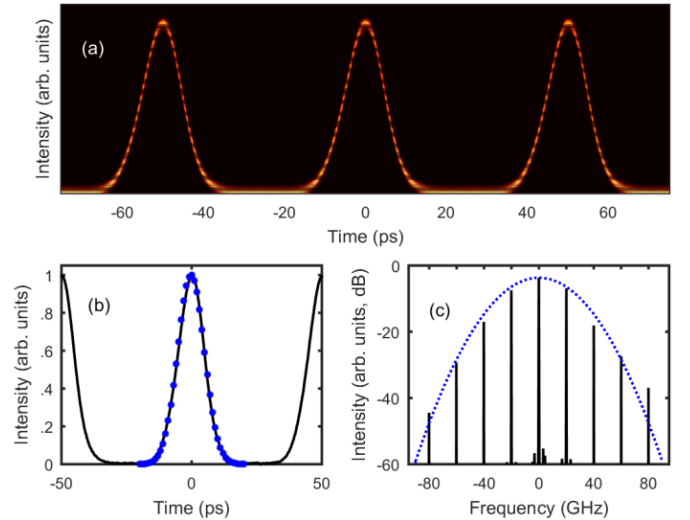

Fig. 5. Pulse properties achieved for $A_m = 1.1$ rad and a repetition rate $f_m = 20$ GHz. (a) Pulse train recorded in persistent mode. Temporal and spectral intensity profiles (panel b and c, respectively). The experimental results (solid black lines) are compared to a fit by a Fourier-transform Gaussian waveform (blue full circles and blue dotted line).

Finally, we have also tested the capacity of our experimental setup to generate higher repetition rate pulse trains. Results obtained for $f_m$ = 30 and 40 GHz are plotted in Fig. 6, panels (a) and (b) respectively. Temporal durations of 7.5 and 6.8 ps have been recorded, leading to duty cycles of 0.22 and 0.27. The slightly degraded results at 40 GHz are mainly explained by the bandwidth limitations of our phase modulator that leads to a decrease of the amplitude of the phase modulation that we can imprint in practice at high frequencies ($A_m$ is indeed limited to 1 rad at 40 GHz). Results plotted on a logarithmic scale (panels (2) of Fig. 6) outline that the floor level is close to -20 dB, confirming the excellent extinction ratio achieved at both repetition rates. Once again, the temporal intensity profiles exhibit an excellent pulse quality: no sidelobes are observed and the profiles are closely adjusted by a Gaussian waveform (blue full circles in panels a2 and b2).

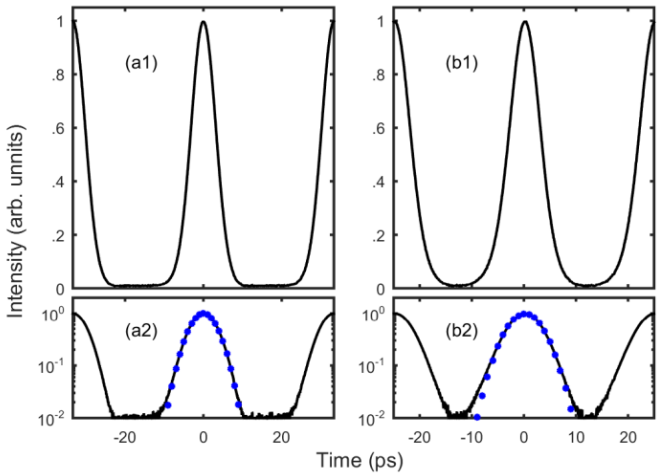

Fig. 6. Temporal intensity profiles obtained after triangular phase compensation. Experimental results obtained at $f_m$ = 30 and 40 GHz are plotted in panels a and b for $A_m$ = 1.1 and 1 rad respectively. The results are displayed on a linear and logarithmic scale (panels 1 and 2). The blue full circles correspond to a fit by a Fourier-transform Gaussian pulse.

## IV. Conclusion

To conclude, we have theoretically and experimentally demonstrated the generation of high-repetition rate trains of ultrafast pulses from the periodic modulation of the phase of a CW laser source. The key ingredient to achieve high-quality pulses is the perfect phase compensation process provided by a simple triangular spectral phase profile. Optimum extinction ratio is obtained for a phase modulation of 1.1 rad for which the resulting pulses are extremely close to a Fourier-transform limited Gaussian shape. Experimental demonstrations carried out at repetition rates ranging from 10 GHz up to 40 GHz fully confirm the theoretical analysis as well as the versatility of this novel technic. Since the present architecture does not involve long pieces of fibers or nonlinear effects, highly stable operation over hours has been recorded without any significant deviations. Combined with additional nonlinear reshaping techniques, it is expected that the duty-cycle can be further reduced [4, 19, 20] and pulses of only a few picoseconds should be achieved.